\documentclass[aps,prl,twocolumn,groupedaddress,showpacs,amsmath]{revtex4}
\usepackage{ams, amsmath, graphicx, epsfig, textcomp, wasysym,html,hyperref}

\begin{document}
\title{Electronic Radio-Frequency Refrigerator}
\author{S.~Kafanov$^{1}\footnote{Electronic address: sergey.kafanov@ltl.tkk.fi}$, A.~Kemppinen$^{2}$, Yu.\,A.~Pashkin$^{3\footnote{On leave from P.N. Lebedev Physical Institute of the Russian Academy of Sciences, Moscow 119991, Russia}}$, M.~Meschke$^1$, J.\,S.~Tsai$^3$ and  J.\,P.~Pekola$^{1}$}
\affiliation{$^1$Low Temperature Laboratory, Helsinki University of Technology, P.O. Box 3500, 02015 TKK, Finland.}
\affiliation{$^2$Centre for Metrology and Accreditation (MIKES), P.O. Box 9, 02151 Espoo, Finland.}
\affiliation{$^3$NEC Nano Electronics Research Laboratories and RIKEN Advanced Science Institute, 34 Miyukigaoka, Tsukuba, Ibaraki 305-8501, Japan.}
\date{\today}


\begin{abstract}
We demonstrate experimentally that a hybrid single-electron transistor with superconducting leads and a normal-metal island can be refrigerated by an alternating voltage applied to the gate electrode. The simultaneous measurement of the dc current induced by the rf gate through the device at a small bias voltage serves as an in-situ thermometer. 
\end{abstract}

\maketitle
Local cooling has become an interesting topic as nanodevices are getting more diverse. Mesoscopic electron systems \cite{ApplPhysLett.65.3123.Nahum, ApplPhysLett.68.1996.Leivo, ApplPhysLett.86.173508.Clark, RevModPhys.78.217.Giazotto, PhysRevLett.99.047004.Rajauria}, superconducting qubits \cite{Science.314.1589.Valenzuela, NaturePhys.4.612.Ploeg} and nanomechanical oscillators \cite{Nature.443.193.Naik, NaturePhys.4.415.Schliesser} are among the systems of interest in this respect. The electron cooler holds the promise in applications, for instance in spaceborne radioastronomy, where it would present an easy-to-use, light-weight solution for noise reduction, with the further benefit of saving energy. In all the realizations until today the electronic refrigerator was operated by a dc bias voltage. Single-electron Coulomb blockade opens, however, a way to  manipulate heat flow on the level of individual electrons \cite{PhysRevLett.99.027203.Saira}, and to synchronize the refrigerator operation to an external frequency of the ac drive, as was predicted in \cite{PhysRevLett.98.037201.Pekola}. Although the ac operation may not produce more efficient refrigeration than the devices with a constant bias \cite{PhysRevLett.98.037201.Pekola, PhysRevB.77.104517.Kopnin} the former one has a number of important benefits: (\textit{i}) in some instances ac operation is the only available operation mode, (\textit{ii}) non-galvanic continuous drive becomes possible and (\textit{iii}) by ac drive one can produce a thermodynamic refrigeration cycle with electrons as the medium. In this Letter we demonstrate a device, the hybrid single-electron turnstile, which makes use of all the features (\textit{i})\,-\,(\textit{iii}), and whose operating temperature can be lowered by almost a factor of two by the ac drive at the gate.

\begin{figure}
\centering\epsfig{figure=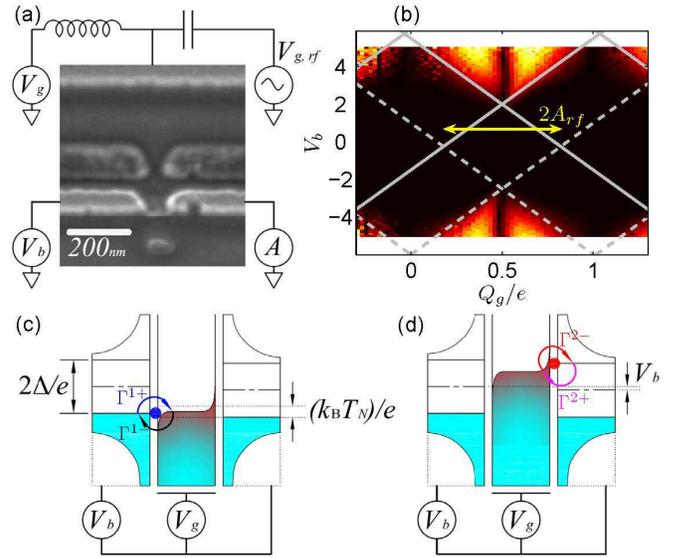,width=\linewidth}
\caption{{\label{Fig1}}(a) Scanning electron micrograph of a typical rf refrigerator with a sketch of its measurement circuity. The darker areas are $\mathrm{AuPd}$ (normal metal), and the brighter ones $\mathrm{Al}$ (superconductor). (b) Measured transconductance $|dI/dV_g|$ of one of the samples as a function of normalized bias voltage $eV_b/\Delta$ and $Q_g=V_gC_g$. The black areas are the stability regions of charge states, where the conductance is negligible. They are limited by the thresholds for tunneling in the direction preferred by positive bias (solid lines) and negative bias (dashed lines). The rf gate signal alternates along the double arrow line. (c, d) The operation principle of the rf refrigerator. There is a normal-metal island between the superconducting bias leads with an energy gap $\Delta$. The dc bias $V_b<2\Delta/e$ defines the preferred direction for tunneling, which is needed for thermometry. The cooling mechanism is based on the periodic single-electron tunneling to and from the normal-metal island driven by the periodic variation of the island potential.}
\end{figure}

The hybrid single-electron transistor (SET) has been intensively studied during the past few years to produce quantized current for metrological applications \cite{NaturePhys.4.120.Pekola, PhysRevLett.101.066801.Averin, ApplPhysLett.94.172108.Kemppinen, arXiv.0803.1563v2.Kemppinen, arXiv.0905.3402v1.Lotkhov}. The rf refrigerator is based on the very same device concept: it is composed of superconducting source and drain leads tunnel coupled to a very small normal-metal island in the Coulomb blockade regime (see Fig.\,\ref{Fig1}(a)).
A small bias voltage applied over the SET defines a preferable direction for single-electron tunneling. However, for the bias voltages $|V|<2\Delta/e$, the dc current through the whole structure is strongly suppressed, due to the superconducting energy gap in the leads. The situation becomes different when a periodic variation of the gate charge of amplitude $A_\mathrm{rf}$ drives the transistor between the stability regions corresponding to two adjacent island charge states (see Fig.\,\ref{Fig1}(b)). Drive transfers a single electron through the turnstile in each cycle, and as a result creates a detectable dc current proportional to the driving frequency ($I\propto f$). The process is associated with heat transport from the island into the bias leads, which is the main topic of the present Letter (see Fig.\,\ref{Fig1}(c, d), for the principle).

The quantitative analysis of the rf refrigerator operation is based on the orthodox theory, where the electron transport is considered as a sequence of instantaneous tunneling events \cite{JETP.62.623.Kulik, Averin}, under the assumption, that the tunneling electrons do not exchange energy with the environment \cite{Grabert.Devoret}. In the quasi-equilibrium limit \cite{RevModPhys.78.217.Giazotto}, the electron energy distribution in the island and in the leads is given by the Fermi-Dirac distribution $f_{N(S)}(\varepsilon)$ with temperature $T_{N(S)}$. In general these temperatures are different from each other and from that of the cryostat, $T_0$. Due to the large volume of the bias leads and the tiny heat flux, we assume that electrons in the leads are well thermalized with lattice phonons ($T_S=T_0$). 

The tunneling rates $\Gamma_{j}^{\pm}$ of electrons tunneling to $(+)$ and from the island $(-)$ through junction with $n$ excess electrons on the island are given by the standard expressions
\begin{equation}
\Gamma_{j,\,n}^{\pm}=\frac{1}{e^2R_{j}}\int n_{S}(\varepsilon)f_{S}(\varepsilon) \left[1-f_{N}(\varepsilon-\delta\mathcal{E}^{\pm}_{j,\,n})\right]d\varepsilon,
\end{equation}
with
\begin{equation}
\delta \mathcal{E}^{\pm}_{j,\,n}=\frac{e^2}{C_\Sigma}\left(n\pm \frac{1}{2}+\frac{V_{g}C_g}{e}\right)+\left(-1\right)^{j}e\frac{C_1C_2}{C_{j}C_\Sigma}V_{b} 
\end{equation}
where $C_{j}$, $R_{j}$ are the capacitance and the resistance of the tunnel junction $j=1,\,2$ and $C_\Sigma=C_1+C_2+C_{g}+C_\mathrm{env}$ is the total capacitance of the island, which includes the capacitance to the gate $C_g$ and that to the environment $C_\mathrm{env}$. The density of states (DOS) in the superconductor is denoted by $n_S(\varepsilon)$. The energy gain $\delta\mathcal{E}_{j,\,n}^{\pm}$ is the decrease of Gibbs energy of the system due to the corresponding tunneling event. The dynamics of electron tunneling through this device is given by the standard master equation for the probability $\sigma_{n,\,t}$ to have $n$ excess electrons on the island \cite{Averin}.
\begin{figure}
\centering\epsfig{figure=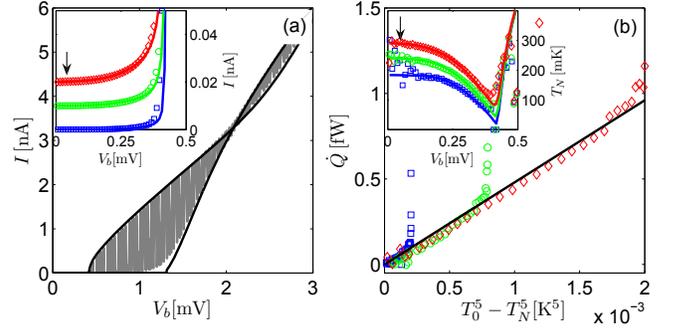,width=\linewidth}
\caption{{\label{Fig2}} (a) Measured (gray) and calculated (black) dc-IV curves. The dc gate voltage was swept during the measurements and hence the envelopes correspond to the expected IV curves with gate open and closed, respectively. The inset shows the measured (open symbols) and calculated (lines) dc-IV curves of the rf refrigerator in the gate-open state at temperatures of $290\,\mathrm{mK}$ (diamonds), $240\,\mathrm{mK}$ (circles) and $186\,\mathrm{mK}$ (squares) from top to bottom. For clarity, all curves are vertically shifted from each other by  $10\,\mathrm{pA}$. The extracted electron temperature (open symbols) of the normal-metal island along these IV curves, and results of the numerical calculations (lines) are shown in the inset of (b). The extracted cooling power vs.~$T_0^5-T_N^5$ for these electron temperatures (open symbols), and a linear fit to the data according to Eq.\,(\ref{Power_balance}) (line) are shown in the main frame of (b). Arrows in the two insets show the bias point for the rf refrigeration experiments.}
\end{figure}

Non-ideality of the superconducting leads can be taken into account assuming a  finite quasiparticle density of states $\gamma$ inside the BCS superconducting gap, \textit{e.g.}, due to inelastic electron scattering in the superconductor  \cite{PhysRevLett.53.2437.Dynes, Kopnin} or by inverse proximity effect due to nearby normal-metals. We model this smeared DOS: $n_{S}(\varepsilon)=\left|\Re\{(\varepsilon-i\Delta\gamma)/
\sqrt{(\varepsilon-i\Delta\gamma)^2-\Delta^2}\}\right|$. Typical experimental value for the effective smearing parameter $\gamma$ for the aluminum thin films near the tunnel junction is $\sim 10^{-4}$ \cite{ApplPhysLett.94.172108.Kemppinen, arXiv.0803.1563v2.Kemppinen}.

Heat transport through the junctions associated to each electron tunneling process is given by
\begin{equation}
\dot{Q}_{j,\,n}^{\pm}=\frac{1}{e^2R_{j}}\int(\varepsilon-\delta\mathcal{E}^{\pm}_{j,\,n}) n_{S}(\varepsilon)f_{S}(\varepsilon) \left[1-f_{N}(\varepsilon-\delta\mathcal{E}_{j,\,n}^{\pm})\right]d\varepsilon.
\end{equation}

The charge current and the cooling power of the rf refrigerator are then given by averaging the corresponding quantities over an operation cycle:
\begin{equation}
I=(-1)^{j}ef\int_{0}^{f^{-1}}\sum_{n} \left(\Gamma_{j,\,n}^{-}-\Gamma_{j,\,n}^{+}\right)\sigma_{n,\,t}dt
\end{equation}
and
\begin{equation}
\dot{Q}=f\int_0^{f^{-1}} \sum_{j,\,n}(\dot{Q}_{n}^{-}-\dot{Q}_{n}^{+})\sigma_{n,\,t}dt.
\end{equation}

The cooling power is counterbalanced by the heat loads from the relaxation processes. The load from electromagnetic coupling to the environment, \textit{i.e.}, electron-photon relaxation, can be ignored due to poor matching between the island and environment \cite{PhysRevLett.93.045901.Schmidt, Nature.444.187.Meschke}. The heat load by electron-phonon interaction dominates in our experiment; the corresponding power is given by \cite{PhysRevLett.55.422.Roukes}  
\begin{equation}
\label{Power_balance}
P_{el-ph}=\Sigma\mathcal{V}(T_{N}^{5}-T_{0}^{5})
\end{equation}
where $\Sigma$ is the electron-phonon coupling constant of the normal-metal and  $\mathcal{V}$ is the island volume. The mean temperature of the island $T_N$ is obtained from $\dot{Q}=P_{el-ph}$. In order to cool the island, the frequency $f$ has to be high enough to prevent full relaxation toward lattice temperature, $\tau_{el-ph}^{-1}\ll f$. On the other hand to secure the quasiequilibrium state of the electron gas we require $f\ll \tau_{el-el}^{-1}$.

The samples were fabricated by electron beam lithography and shadow deposition technique \cite{PhysRevLett.59.109.Fulton, ApplPhysLett.76.2256.Pashkin}, and they were measured in a dilution refrigerator with a base temperature of $40\,\mathrm{mK}$. For the characterization of the rf refrigerator, we measured the IV curves of the device at the base temperature of the cryostat, with simultaneous fast ramping of the gate voltage  (see Fig.\,\ref{Fig2}(a)). The solid lines are the calculated IV curves for the two extreme gate positions: gate-open $Q_g=V_gC_g=e/2$ and gate closed $Q_g=0$. From these fits we get the following parameters of the sample: asymptotic resistance $R_{\infty}=R_1+R_2=315\,\mathrm{k\Omega}$; charging energy $E_c=e^2/(2C_\Sigma)=7\,\mathrm{K}$; superconducting energy gap $\Delta=210\,\mathrm{\mu eV}$; gap smearing parameter $\gamma=2\left.dI/dV\right|_{V=0}R_{\infty}=9.4\times 10^{-5}$.
In order to obtain the value of $\Sigma$, we measured and fitted the IV characteristics in the subgap region at the gate-open state at different cryostat temperatures, see the inset of Fig.\,\ref{Fig2}(a). The corresponding electron temperatures extracted from the fitting are shown in the inset of Fig.\,\ref{Fig2}(b). In the gate-open state, the turnstile functions similarly to a regular SINIS cooler \cite{ApplPhysLett.68.1996.Leivo, PhysRevLett.99.027203.Saira}, with a maximum cooling power at the bias voltage $V_b\simeq\pm 2\Delta/e$. At higher bias voltages, the turnstile operates in the regime where the temperature rapidly increases with bias voltage.
Figure\,\ref{Fig2}(b) presents the extracted cooling power $\dot{Q}$ vs.~$T_0^5-T_N^5$ matched with the heat load from the electron-phonon relaxation. Using the dimensions of the island, $\mathcal{V}=30\times 50 \times 80\,\mathrm{nm}^3$, we then obtain $\Sigma=4\times 10^9\,\mathrm{WK^{-5}m^{-3}}$ from the linear fit of the data, which is in good agreement with the values obtained for the same Au-Pd alloy in another experiment \cite{PhysRevLett.Timoveev}.
\begin{figure}
\centering\epsfig{figure=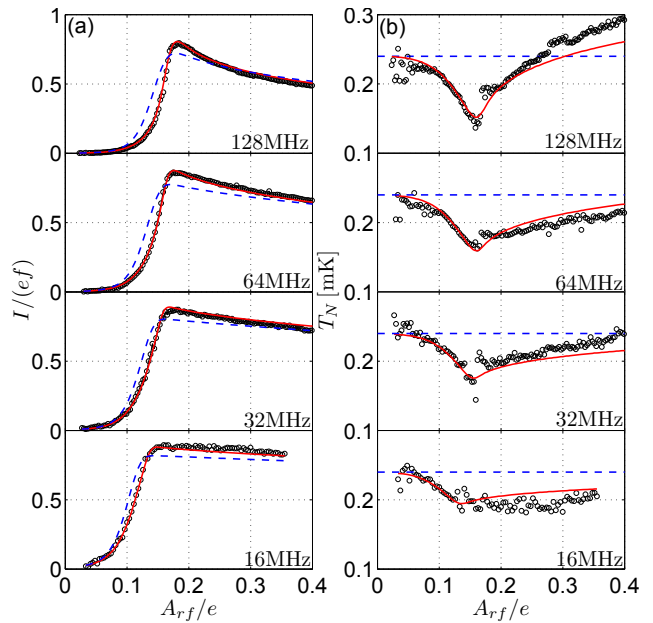,width=\linewidth}
\caption{{\label{Fig3}}(a) Measured current $I$ (circles), scaled by $ef$ vs.~the rf amplitude at different operation frequencies $f$. The data were acquired at $T_0=240\,\mathrm{mK}$ and gate-open state. The rf refrigerator is biased with a low bias voltage $V_b=50\,\mathrm{\mu V}\simeq 0.25\Delta/e$, where dc cooling effect is negligible, see Fig.\,\ref{Fig2}(b). The corresponding calculated current for the constant island temperature ($T_{N}=T_0$) (dashed lines) and for the amplitude dependent temperature of the normal-metal island (solid line), obtained from the power balance equation ($\dot{Q}=P_{el-ph}$). (b) Open circles show the rf amplitude dependent electron temperature of the island, obtained from the measured current presented in (a). The temperature obtained from the simulations is shown by the continuous lines. Cryostat temperature is shown by the dashed horizontal lines.}
\end{figure}

For the demonstration of rf refrigeration, we measured the charge current though the device at different operation frequencies ($f=2^k\,\mathrm{MHz},\,k=1\dots 7$), and at different bath temperatures $100\,\mathrm{mK}\apprle T_0\apprle 500\,\mathrm{mK}$. In order to distinguish between the ordinary dc cooling and rf cooling, we biased the turnstile at a low voltage $V_b=50\,\mathrm{\mu V}\simeq 0.25\Delta/e$, where dc cooling is small. This bias is indicated by the arrows in the insets of Fig.\,\ref{Fig2}. Generally, the dc bias is not needed for rf cooling, but it makes in-situ thermometry possible.

The measured current in the gate-open state vs.~the rf amplitude at different frequencies is shown in Fig.\,\ref{Fig3}(a). The cryostat temperature was $T_0=240\,\mathrm{mK}$ in this case. With a small bias voltage $V_b$ applied, the rates of tunneling in the forward and backward directions differ by a factor of $\sim\exp\left(-eV_b/(k_{\mathrm{B}}T_{N})\right)$ \cite{NaturePhys.4.120.Pekola}. Thus, measuring the dc current $I$ through the device serves as a thermometer of the island. By using parameters of the cooler obtained from the dc measurements, and taking into account the balance between the cooling power and the heat flow due to the electron-phonon relaxation, we have calculated the corresponding current $I$ as a function of rf amplitude; the simulation results are shown by a continuous line in Fig.\,\ref{Fig3}(a). As a reference we also show (dashed lines), the corresponding curves calculated for fixed temperature ($T_N=T_0$). We should mentioned that we used $Q_g$ as a fitting parameter in Fig.\,\ref{Fig3}. Good agreement between the experiment and simulations with non-constant $T_N$ allows us to extract the temperature of the island. Figure\,\ref{Fig3}(b) shows the mean temperature $T_N$ thus obtained (open symbols), and the corresponding predicted temperature (continuous lines) from the numerical simulations with the independently determined parameters. We note that the instantaneous electron temperature in the rf refrigerator island is expected to fluctuate around its mean value $T_N$, due to fundamental principles of thermodynamics. These fluctuations are inversely proportional to the volume of the island,  $\langle \delta T_N^2\rangle=k_{\mathrm{B}}T_N^2/C_\mathrm{el}\propto T_N/\mathcal{V}$, where $C_\mathrm{{el}}$ is the heat capacity of the electron gas in the island \cite{Landau}. For our samples, with a very small island, we obtain $\langle \delta T_N^2\rangle^{1/2}\sim 10\,\mathrm{mK}$ at $T_N\simeq 200\,\mathrm{mK}$. 
\begin{figure}
\centering\epsfig{figure=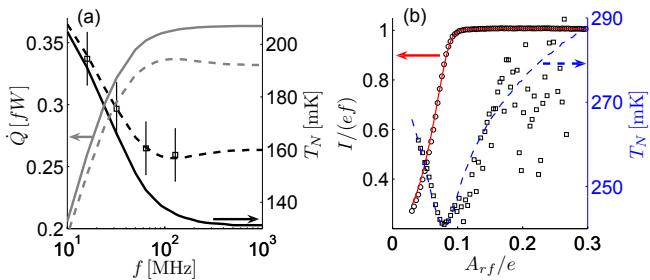,width=\linewidth}
\caption{{\label{Fig4}} (a) The calculated maximum of the cooling power (gray lines) and the lowest temperature (black lines) of the normal-metal island vs.~the rf frequency, for two different gate charges, $Q_g=0.5e$ (solid lines) and $Q_g=0.48e$ (dashed lines); temperature and the bias voltage are the same as those  used in Fig.\,\ref{Fig3}. Open squares show the minimum temperatures obtained in the experiment, see Fig.\,\ref{Fig3}(b). The error bars are obtained based on data in Fig.\,\ref{Fig3}(b). (b) Measured (circles), and calculated (continuous line) rf induced dc current through the device at $f=64\,\mathrm{MHz}$. Open squares show the extracted electron temperature of the island; dashed line shows the expected temperature behavior based on the simulations. The data was acquired in the gate-open state; the bias was set to the optimum position for pumping \cite{NaturePhys.4.120.Pekola, PhysRevLett.101.066801.Averin, ApplPhysLett.94.172108.Kemppinen} $V_b=200\,\mu\mathrm{V}\simeq\Delta/e$. The base temperature was $T_0=300\,\mathrm{mK}$, which is available in a $^3\mathrm{He}$ cryostat.}
\end{figure}

Figure\,\ref{Fig4}(a) shows the calculated cooling power (gray lines) and the corresponding minimum temperature of the island (black lines) for two different dc gate charges ($Q_g=0.5e$ and $0.48e$). The highest cooling power is achieved exactly in the gate-open state. The cooling power decreases rapidly, even for small offsets from this position, because cooling is not any more optimized for tunneling through both junctions. Therefore, background charge fluctuations reduce the cooling power of the refrigerator, and affect its temperature. However, the cooling power increases with operation frequency. For the frequencies lower than the characteristic electron-phonon relaxation rate, the electron temperature is close to the lattice temperature. At higher frequencies, the cooling power rises monotonically and eventually saturates due to the finite $R_{\infty}C_{\Sigma}$ time constant of the device. Because of the small drive amplitude of the rf refrigerator the frequency dependence of the cooling power does not turn into  heating at high frequencies, which, on the other hand, is predicted for multi-electron cycles \cite{PhysRevLett.98.037201.Pekola}.  

The rf refrigeration plays an important role in the development of the current standard based on the hybrid turnstile. This effect allows one to cool down the island also in the metrologically interesting range of the operation parameters. The experimental pumping curve measured at $T_0=300\,\mathrm{mK}$ with a plateau at $I=ef$ and the extracted electron temperature at $f=64\,\mathrm{MHz}$ are shown in Fig.\,\ref{Fig4}(b). Here, the turnstile is in the gate-open state and biased at the optimum bias point for pumping, $V_b\simeq\Delta/e$.

In conclusion, we have experimentally demonstrated rf refrigeration using a single-electron transistor with superconducting leads and a normal-metal island, by applying an rf signal to the gate electrode. The cooling power rises monotonically with operation frequency until it saturates. In practice the demonstrated rf cooling effect may be useful \textit{e.g.}, in the development of a standard for electric current.

This work was partially supported by the Academy of Finland, Japan Science and Technology Agency through the CREST Project, the European Community's Seventh Framework Program under Grant Agreement No.\,218783 (SCOPE), the NanoSciERA project "NanoFridge" and EURAMET joint research project REUNIAM, the Technology Industries of Finland Centennial Foundation.

\end{document}